\documentstyle[12pt,epsfig,rotating]{article}
\newcommand{\pbarn}{\hbox {pb}}
\newcommand{\lumun}{{\hbox {pb}^{-1}}{\hbox {yr}^{-1}}}
\newcommand{\hc}{\hbox {h.c.}}

\newcommand{\im}{\hbox {Im}}

\newcommand{\gev}{\hbox {GeV}}
\newcommand{\etal}{\hbox{et al.}}
\newcommand{\prdj}[1]{{ \it Phys.~Rev.}~{\bf D{#1}}}
\newcommand{\prlj}[1]{{ \it Phys.~Rev.~Lett.}~{\bf {#1}}}
\newcommand{\plbj}[1]{{ \it Phys.~Lett.}~{\bf {#1B}}}
\newcommand{\npbj}[1]{{ \it Nucl.~Phys.}~{\bf B{#1}}}
\newcommand{\ptpj}[1]{{ \it Prog.~Theor.~Phys.}~{\bf {#1}}}
\newcommand{\zfpj}[1]{{ \it Z.~Phys.}~{\bf C{#1}}}
\newcommand{\ttbar}{\mbox{$t\bar{t}$}}

\newcommand{\tdec} {\mbox{$t\rightarrow W^+ b$}}
\newcommand{\tbdec} {\mbox{$\bar{t} \rightarrow W^- \bar{b}$}}
\newcommand{\ttprod} {\mbox{$e^-e^+ \rightarrow \ttbar$}}
\newcommand{\sstop}{\tilde t}
\newcommand{\tbar} {\mbox{$\bar{t}$}}
\newcommand{\bbar} {\mbox{$\bar{b}$}}
\newcommand{\nubar} {\mbox{$\bar{\nu}$}}
\newcommand{\wpp} {\mbox{$W^+$}}
\newcommand{\wm} {\mbox{$W^-$}}
\newcommand{\wpdec} {\mbox{$\wpp \rightarrow  \ell^+ \nu$}}
\newcommand{\wmdec} {\mbox{$\wm \rightarrow   \ell^- \nubar$}}
\newcommand{\wphel}{\mbox{$\lambda_{W^+}$}}
\newcommand{\wmhel}{\mbox{$\lambda_{W^-}$}}
\newcommand{\bhel} {\mbox{$h_{b}$}}
\newcommand{\bbhel} {\mbox{$h_{\bar b}$}}
\newcommand{\thel} {\mbox{$h_{t}$}}
\newcommand{\tbhel} {\mbox{$h_{\bar t}$}}
\newcommand{\lphel} {\mbox{$h_{\ell^+}$}}
\newcommand{\lmhel} {\mbox{$h_{\ell^-}$}}
\newcommand{\nuhel} {\mbox{$h_{\nu}$}}
\newcommand{\nubhel} {\mbox{$h_{\bar \nu}$}}
\newcommand{\tdechel} {\mbox{$\thel,\wphel,\bhel$}}
\newcommand{\tbdechel} {\mbox{$\tbhel,\wmhel,\bbhel$}}
\newcommand{\wpdechel} {\mbox{$\wphel,\lphel,\nuhel$}}
\newcommand{\wmdechel} {\mbox{$\wmhel,\lmhel,\nubhel$}}
\newcommand{\astop} {\mbox{${\cal A}^t$}}
\newcommand{\astopb} {\mbox{${\cal A}^{\bar t}$}}
\newcommand{\as} {\mbox{${\cal A}$}}
\newcommand{\st} {\mbox{$s_t$}}
\newcommand{\stb} {\mbox{$s_{\bar t}$}}
\newcommand{\swp} {\mbox{$s_{W^+}$}}
\newcommand{\swm} {\mbox{$s_{W^-}$}}
\newcommand{\ml} {\mbox{$m_\lambda^2$}}
\newcommand{\mst}  {\mbox{$m_{{\tilde{t}}  }^2$}}

\newcommand{\msb} {\mbox{$m_{\tilde{b}}^2$}}
 
\begin{document}
\centerline{hep-ph/9306322
\hfill
CERN-TH.6917/93}
\vspace*{2cm}
\begin{center}
{\large{\bf SUSY-Induced CP Violation in {\boldmath $t$} Decays
at {\boldmath $e^- e^+$} Colliders} \\
\vspace*{2cm}
Bohdan Grz\c{a}dkowski\footnote{E-mail:
{\tt bohdang@cernvm.}\\
\noindent Address after 1 October 1993:
        Institute for Theoretical Physics,
        University of Warsaw,
        Ho\.{z}a 69, PL-00-681 Warsaw, Poland. }\\
\vspace*{.5cm}
Wai-Yee Keung\footnote{E-mail: {\tt keung@cernvm.}\\
\noindent Address after 1 August 1993:
        Physics Department, M/C 273
        University of Illinois at Chicago,
        IL 60607--7059,
        USA. }\\
\vspace*{1cm}
       Theory Division, CERN, CH-1211 Geneva 23, Switzerland }\\
\vspace*{3cm}
 
{\bf Abstract} \\
\end{center}
CP violation in the decays $t \to \ell^+\nu b$ and $\bar t \to
\ell^-\bar\nu \bar b$ from the production process $e^-e^+ \rightarrow \ttbar$
is discussed. Since the asymmetry proposed as a measure of CP
violation  vanishes even at the one-loop level in the Standard Model
(SM), it may be a useful tool to search for sources of CP
violation outside  of the SM. As an illustration the asymmetry has been
computed within  supersymmetric extensions of the SM. Prospects for its
measurement at future high-energy linear colliders are discussed.
\vspace*{2.0cm}
\begin{flushleft}
\parbox{5in}
{Published in Phys. Lett. {\bf B316} (1993) 137\\
CERN-TH.6917/93\\
June 1993}
\end{flushleft}
\setcounter{page}{0}
\thispagestyle{empty}
\newpage
 
\section{Introduction}
 
Collider physics can prove to be an important area, complementary to the
low-energy kaon and B-meson physics, in searching  for
CP violation~\cite{cpcol,werner}. In particular, future linear high-energy
$e^-e^+$ colliders can provide very useful laboratories to study CP
violation in the top-quark system. The top quark being very heavy can
offer a few relevant advantages for the  study of CP violation:
\begin{itemize}
\item{If $m_t>130$ GeV, it would decay before it can form a bound
 state~\cite{bigi};
therefore the perturbative description is much more reliable.}
\item{For the same reason, spin effects would not be
diluted by hadronization effects --- which provides a very useful tool
in searching for CP-violating effects.}
\item{Again, because of its large mass, the top quark properties are
sensitive to interactions mediated by Higgs bosons~\cite{lee}.}
\item{The Kobayashi-Maskawa~\cite{km} mechanism of CP violation
is strongly suppressed for the top quark since its mixing with other
generations is very weak; therefore it is sensitive to non-conventional sources
 of CP violation.}
\end{itemize}
Simultaneously one should bear in mind that in spite of spectacular successes
obtained in experimental high-energy
physics (e.g.    precision tests of the Standard Model)
the origin of CP violation is still a mystery.
The standard theory of Kobayashi and Maskawa~\cite{km}
provides explicit CP violation through phases of Yukawa couplings;
however, many other attractive schemes of CP violation
may emerge in extensions of the Standard Model (SM).
Those unconventional sources of CP violation can be
tested in $\ttbar$ production at future linear $e^-e^+$ colliders.
For instance, CP violation induced by the neutral scalar
sector              in the two-Higgs-doublet standard
model~\cite{lee} (2HDM) may lead to observable effects~\cite{keung}
in $\ttprod$, provided sufficient luminosity can be obtained.
The supersymmetric
extension of the Standard Model (SSM) can also produce CP violation
at the one-loop
level both in $\ttbar$ production and in its
decay~\cite{bg,susywerner} at $e^-e^+$ colliders.
One can also imagine some unknown
high-scale theory, which induces effective, low-energy, CP-violating
interactions.
Here, we will generalize the method of observation of CP violation, proposed
 originally~\cite{gg} for $W$--$g$ fusion at a {\it pp} collider, to  the $e^-e^+$  environment,
and calculate the CP-violating asymmetry within
a supersymmetric extension of the Standard Model (SSM). We will emphasize
that the asymmetry we define is (at least) at the leading order of
perturbation expansion a measure of CP violation in the top quark
decay and that it is not affected by any CP violation taking place at \ttbar\
 production.
 
\section{The Strategy}
 
We will consider production of \ttbar\ pairs at future high-energy
linear colliders. We assume that the production mechanism can be described by the
 $\gamma$ and $Z$ exchange where CP violation may enter at \ttbar\ production
 vertex. The production is then followed by the decays $\tdec$ and $\tbdec$, and
 eventually by $\wpdec$ and $\wmdec$.
Again, CP violation may be present at \ttbar\ decays; however we assume that
 \wpp\ and \wm\ decay conventionally, i.e. without CP violation.
 
In order to describe $t$ and $\bar{t}$ decays we will use the most general
 4-form factor \\ parametrization of $\tdec$ and $\tbdec$
decay vertices :
\begin{eqnarray}
\Gamma^\mu & = & \frac{-igV^{KM}_{tb}}{\surd 2}\bar{u}(p_b)\left[
\gamma^\mu (f^L_1 P_L+f^R_1 P_R)-\frac{i\sigma^{\mu \nu } k_\nu }{m_W}
(f^L_2 P_L+f^R_2 P_R)\right] u(p_t)
\label{decays}
,\\
\bar{\Gamma}^\mu & = & \frac{-ig{V^{KM}_{tb}}^\ast}
{\surd 2}\bar{v}(p_{\bar{t}})\left[
\gamma^\mu (\bar{f}^L_1 P_L+\bar{f}^R_1 P_R)-
\frac{i\sigma^{\mu \nu } k_\nu }{m_W}
(\bar{f}^L_2 P_L+\bar{f}^R_2 P_R)\right] v(p_{\bar{b}}),
\end{eqnarray}
where $P_{R/L}$ are projection operators, $k$ is the
$W$ momentum, $V^{KM}$ is the Kobayashi--Maskawa matrix
and $g$ is the $SU(2)$ gauge coupling constant.
Since $W$ decays into massless fermions, two other, in principle present, form
 factors do not contribute.
 
It is easy to show that~\cite{werner,gg} :
\begin{equation}
f^{L,R}_1=\pm \bar{f}^{L,R}_1,\;\;\;\;\;\;f^{L,R}_2=\pm \bar{f}^{R,L}_2,
\label{cptrans}
\end{equation}
where the upper (lower) signs are those for contributions induced by
CP-conserving \\ (-~violating) interactions.
 
The general form of the phase-space element for the process
$e^-e^+ \rightarrow \ttbar \rightarrow \wpp \wm b \bbar \rightarrow
\ell^+ \ell^- \nu \nubar b \bbar$,
\begin{equation}	
d\Phi=(2\pi)^4 \prod^6_{i=1} \frac{d^3p_i}{(2\pi)^3 2E_i}	
\delta^{(4)}(P_{in}-\sum^6_{i=1}p_i),
\label{phase}
\end{equation}
could be written as:
\begin{eqnarray}
d\Phi&=&(2\pi)^{-4} d\st d\stb d\swp d\swm d\Phi(\ttprod) \times
\nonumber \\
  & &   d\Phi(\tdec) d\Phi(\wpdec) d\Phi(\tbdec) d\Phi(\wmdec),
\label{newphase}
\end{eqnarray}
where $s_x$ denotes the corresponding invariant mass of the decaying particles.
The appropriate matrix element for a given helicity final state reads:
\begin{eqnarray}
\lefteqn{
(\bhel,\bbhel,\lphel,\lmhel,\nuhel,\nubhel)=
D_t(\st)D_{\bar t}(\stb)D_{W^+}(\swp)D_{W^-}(\swm)
\sum_{\thel \tbhel}(\thel,\tbhel) } \\
&&\times \sum_{\wphel}(\tdechel)(\wpdechel)
\sum_{\wmhel}(\tbdechel)(\wmdechel), \nonumber
\label{ampl}
\end{eqnarray}
where $h_x$ denotes the $x$-helicity and the helicities of the initial $e^-e^+$  have
 been omitted in the amplitudes for \ttprod : $(\thel,\tbhel)$;
$D_x$ are the propagators of decaying $t$, $\bar{t}$, \wpp\ and $W^-$. Since we
 are interested in decays of real particles
we will use for $D_x(s_x)$ the Breit-Wigner
parametrization :
\begin{equation}
D_x(s_x)\equiv \frac{1}{s_x-m_x^2+im_x\Gamma_x} \;\; .
\end{equation}
The absolute value of the amplitude squared can be written as :
\begin{eqnarray}
|(\cdots)|^2&=&|D_t(\st)|^2|D_{\bar t}(\stb)|^2
|D_{W^+}(\swp)|^2|D_{W^-}(\swm)|^2
\sum_{\thel \tbhel} \sum_{\thel' \tbhel'} (\thel,\tbhel)(\thel',\tbhel')^
\ast \times \nonumber \\
&&\sum_{\wphel}(\tdechel)(\wpdechel)\sum_{\wphel'}(\thel',\wphel',\bhel)^
\ast(\wphel',\lphel,\nuhel)^\ast \times \\
&&\sum_{\wmhel}(\tbdechel)(\wmdechel)\sum_{\wmhel'}(\tbhel',\wmhel',\bbhel)
^\ast(\wmhel',\lmhel,\nubhel)^\ast. \nonumber
\label{ampsq}
\end{eqnarray}
Within the narrow-width approximation~\cite{zepenfeld} one can easily perform
 phase-space integrations	
over invariant masses of decaying $t$, \tbar, \wpp and \wm since:
\begin{equation}
|D_x(s_x)|^2\simeq \frac{\pi}{m_x \Gamma_x} \delta(s_x-m_x^2).
\end{equation}
We concentrate on the decay products of the top quark and therefore we shall
 integrate over $d\Phi(\tbdec)$ and $d\Phi(\wmdec)$:
\begin{eqnarray}
\lefteqn{
\int d\Phi(\tbdec) d\Phi(\wmdec)
\sum_{\wmhel}(\tbdechel)(\wmdechel)\times}     \\
 &&\sum_{\wmhel'}(\tbhel',\wmhel',\bbhel)^\ast
                (\wmhel',\lmhel,\nubhel)^\ast
=\delta_{\tbhel,\tbhel'} \;2m_t \Gamma(\tbdec) \;2m_W \Gamma(\wmdec),
\nonumber
\end{eqnarray}
where a summation over the final-state helicities has been implicitly assumed.
It will be useful to define normalized (Tr $\rho=1$) top-quark density matrix:
\begin{equation}
\sum_{\tbhel}(\thel,\tbhel)(\thel',\tbhel)^\ast=
\rho_{\thel \thel'}
\sum_{\thel \tbhel}|(\thel,\tbhel)|^2\;\; ,
\label{densmat}
\end{equation}
again summed over initial $e^-e^+$  helicities.
 
The differential cross-section for the total decay sequence could now be
written as :
\begin{eqnarray}
\frac{d\sigma_{tot}}{d\Phi(t   \rightarrow W^+b)\;
                     d\Phi(W^+ \rightarrow \ell^+ \nu)}
 = \frac{B\!R(\tbdec)}{2m_t\Gamma_t}\;
\frac{B\!R(\wmdec)}{2m_W\Gamma_W} d\sigma(\ttprod)
\times \nonumber \\
\sum_{\thel \thel'}\rho_{\thel \thel'}
 \sum_{\wphel}(\tdechel)(\wpdechel)
\left[\sum_{\wphel'}(\thel',\wphel',\bhel)(\wphel',\lphel,\nuhel)\right]^\ast
\;.
\end{eqnarray}
The factor $B\!R(\wmdec)$ will become the branching fraction of $W^-$
into hadrons if we use the more abundant non-leptonic mode.
Now, we are in a position to make use of the asymmetry defined in Ref.~\cite{gg}
by the integration over the total top-quark decay phase space $d\Phi(\tdec)$
and a restricted one for $(\ell^+\nu)$:
\begin{equation}
{\cal A}^t \equiv \frac{N^t}{D^t},
\end{equation}
where
\begin{eqnarray}
N^t &\sim& \int d\Phi(\tdec) \int_1^{-1} d\cos(\theta_{\ell^+})
\left[\int_0^\pi-\int_{-\pi}^0\right] d\phi_{\ell^+}
 \frac{d\sigma_{tot}}{d\Phi(\tdec)\;d\Phi(\wpdec)} \nonumber \\
D^t &\sim& \int d\Phi(\tdec) \int_1^{-1} d\cos(\theta_{\ell^+})
\left[\int_0^\pi+\int_{-\pi}^0\right] d\phi_{\ell^+}
 \frac{d\sigma_{tot}}{d\Phi(\tdec)\;d\Phi(\wpdec)}, \nonumber
\end{eqnarray}
$\theta_{\ell^+}$ and $\phi_{\ell^+}$ are the polar and azimuthal angles
of the $\ell^+$ defined in the $W^+$ rest frame~\cite{gg}.
For the density matrix we shall adopt the following
parametrization~\cite{kane} :
\begin{equation}
\rho^t=\frac{1}{2} \left[
\begin{array}{cc}
1+P_\parallel^t & P_\perp^t e^{i\alpha} \\
P_\perp^t e^{-i\alpha} & 1-P_\parallel^t,
\end{array}
\right]
\end{equation}
where $P_\perp^t$ and $P_\parallel^t$ describe the
polarization of the top quark~\cite{kane} :
\begin{equation}
s^\mu\equiv P_\perp^t(0,\hat{\mbox{\boldmath $s$}}) + \frac{P_\parallel^t}{m_t}
(|\mbox{\boldmath $p$}_t|,E_t\hat{\mbox{\boldmath $p$}}_t).
\end{equation}
In the above equation, $E_t$ and $\hat{\mbox{\boldmath $p$}}_t$ denote the top-quark
 energy and the  direction of its 3-momentum, respectively;
$\hat{\mbox{\boldmath $s$}}$ is a unit 3-vector perpendicular to
 $\hat{\mbox{\boldmath $p$}}_t$ and the angle $\alpha$ specifies the direction
 of $\hat{\mbox{\boldmath $s$}}$. A direct calculation leads to the following result :
\begin{equation}
\astop=h(m_t)\im(f_1^L{f_2^R}^\ast)\frac{d\sigma(\ttprod)
 P_\parallel^t(\theta_{\ttbar})}
{d\sigma(\ttprod)},
\end{equation}
where
$$h(m)\equiv\frac{3\pi}{8}\frac{m^2-m_W^2}{m^2+2m_W^2}.$$
Form factors $f_1^L$ and $f_2^R$ are defined in Eq.~(\ref{decays}).
The tree-level value $f_1^L=1$ will be adopted hereafter.~\footnote{
Notice that $f_2^R$ vanishes at the tree level. It is easy to see that $f_2^R$
is zero in the SM, even at the one-loop level; therefore, \as\ is very sensitive
to non-standard sources of CP violation.}
As we can  see, only parallel-spin components are relevant for our
considerations. Under $\phi_{\ell^+}$ integration all non-diagonal elements
of the density matrix  cancelled.
It is worth-while to remember that the longitudinal polarization
$P_\parallel^t$ is a function of the \ttbar\ production angle $\theta_{\ttbar}$.
 
Imaginary parts of $f_2^R$ may be induced not only by CP-violating interactions,
but also by
final-state CP-conserving interactions like $\gamma$ or $Z$ exchange
between final $b$ and $W^+$.
In order to cancel all such CP-conserving contributions~\cite{gg}
to $\astop$, we define a new
asymmetry $\as$, adding to \astop\ its analogue for \tbar, \astopb\ :
\begin{equation}
\as\equiv\astop + \astopb.
\end{equation}
Let us split CP-conserving (CPC) and CP-violating (CPV) contributions
to $f_2^R$:
$f_2^R={f_2^R}_{CPC}+{f_2^R}_{CPV}$. Using relations,
see Eq.~(\ref{cptrans}), between form factors for top and antitop decays one can
 write \as\ as :
\begin{equation}
\as=-h(m_t)\frac{[\im({f_2^R}_{CPC})(P_\parallel^t + P_\parallel^{\tbar})
                + \im({f_2^R}_{CPV})(P_\parallel^t - P_\parallel^{\tbar})]
                     d\sigma(\ttprod)}
                    {d\sigma(\ttprod)}.
\end{equation}
However, since $(P_\parallel^t+P_\parallel^{\tbar})$ vanishes
at the tree-level approximation to the production mechanism~\cite{kane},
it is seen that \as\ is defined in such a way that
all CP-conserving contributions to \astop\ and \astopb\ cancel
in leading order :
\begin{equation}
\as=-2h(m_t)\im({f_2^R}_{CPV})\frac{P_\parallel^t(\theta_{\ttbar})d\sigma
(\ttprod)}{d\sigma(\ttprod)}.
\label{asymmetry}
\end{equation}
The above shows that our asymmetry is not sensitive to possible CP violation
in the \ttbar\ production mechanism, since the production entered at the tree
 level through $P_\parallel^t$.
The longitudinal polarization is given by the following formula~\cite{kane}:
\begin{equation}
P_\parallel^t(\theta_{\ttbar})=\frac{|(++)|^2+|(+-)|^2-|(-+)|^2-|(--)|^2}
{|(++)|^2+|(+-)|^2+|(-+)|^2+|(--)|^2},
\end{equation}
 
In order to calculate $P_\parallel^t$ we will assume that the production process
 \ttprod\ can be correctly described by $\gamma$ and $Z$ $s$-channel exchange. We
 start by writing down the general form factors of the $Vt\bar{t}$ interaction
 ($V=\gamma, Z$).
The vertex amplitude $ ie \Gamma^V$ for the virtual $V$ can be
parametrized by the following expression:
\begin{equation}
  \Gamma^V_\mu  = c_v^V \gamma_\mu  + c_a^V \gamma_\mu\gamma_5
                + c_d^V i\gamma_5 {{p_t}_\mu-{p_{\bar{t}}}_\mu\over 2 m_t} +
 \cdots.
\label{eqn:form}
\end{equation}
We use the tree-level values for  $c_v$ and $c_a$. They are
\begin{eqnarray}
  c_v^\gamma=&2/3, \;\;\;\;\;\;\;\; c_a^\gamma=0,          \nonumber\\
   c_v^Z=& [1/4 - 2/(3 s_W^2)]/\sqrt{s_W^2(1-s_W^2)}       \\
   c_a^Z=& -1/[4\sqrt{s_W^2(1-s_W^2)}],
\;   \nonumber
\end{eqnarray}
where $s_W\equiv\sin(\theta_W)$.
 
The helicity amplitudes $(h_t,h_{\bar t})$
for the process $e^-e^+\rightarrow t \bar t$ at the scattering
angle $\theta_{\ttbar}$ have been given in the literature~\cite{kane,keung}.
For the initial $e^-e^+$  helicity configuration of $(-+)$, we have
\begin{eqnarray}
(-+)=&e^2[c_v^\gamma+r_L c_v^Z-\beta r_L c_a^Z](1+\cos\theta_{\ttbar})
                                                               \nonumber\\
(+-)=&e^2[c_v^\gamma+r_L c_v^Z+\beta r_L c_a^Z](1-\cos\theta_{\ttbar})
                                                               \nonumber\\
(--)=&e^2[2t(c_v^\gamma+r_L c_v^Z)
           - (i/2) (c_d^\gamma + r_L c_d^Z)\beta /t] \sin\theta_{\ttbar}
\label{eq:amp}
                                                                    \\
(++)=&e^2[2t(c_v^\gamma+r_L c_v^Z)
           + (i/2) (c_d^\gamma + r_L c_d^Z)\beta /t] \sin\theta_{\ttbar}
                                                                 \;.\nonumber
\end{eqnarray}
The dimensionless variables are defined by $t=m_t/\sqrt{s}$,
$z=m_Z/\sqrt{s}$, $\beta^2=1-4t^2$. The $Z$ propagator and its
coupling to the left handed electron gives $-e r_L/s$, with
\begin{equation}
      r_L=( 1/2 - s_W^2)/ [(1-z^2) \sqrt{s_W^2 (1-s_W^2)}] \; .
\end{equation}
Similarly, we obtain formulas for the initial $e^-e^+$ configuration $(+-)$
with $r_L$ replaced by $r_R$,
\begin{equation}
      r_R=-s_W^2 /[(1-z^2) \sqrt{s_W^2 (1-s_W^2)}] \; ,
\end{equation}
and $\cos\theta_{\ttbar}$ by $-\cos\theta_{\ttbar}$, 
and $\sin\theta_{\ttbar}$ by $-\sin\theta_{\ttbar}$
in Eq.~(\ref{eq:amp}).
 
\begin{figure}[hbt]
\begin{center}
\begin{sideways}
\mbox{\epsfig{file=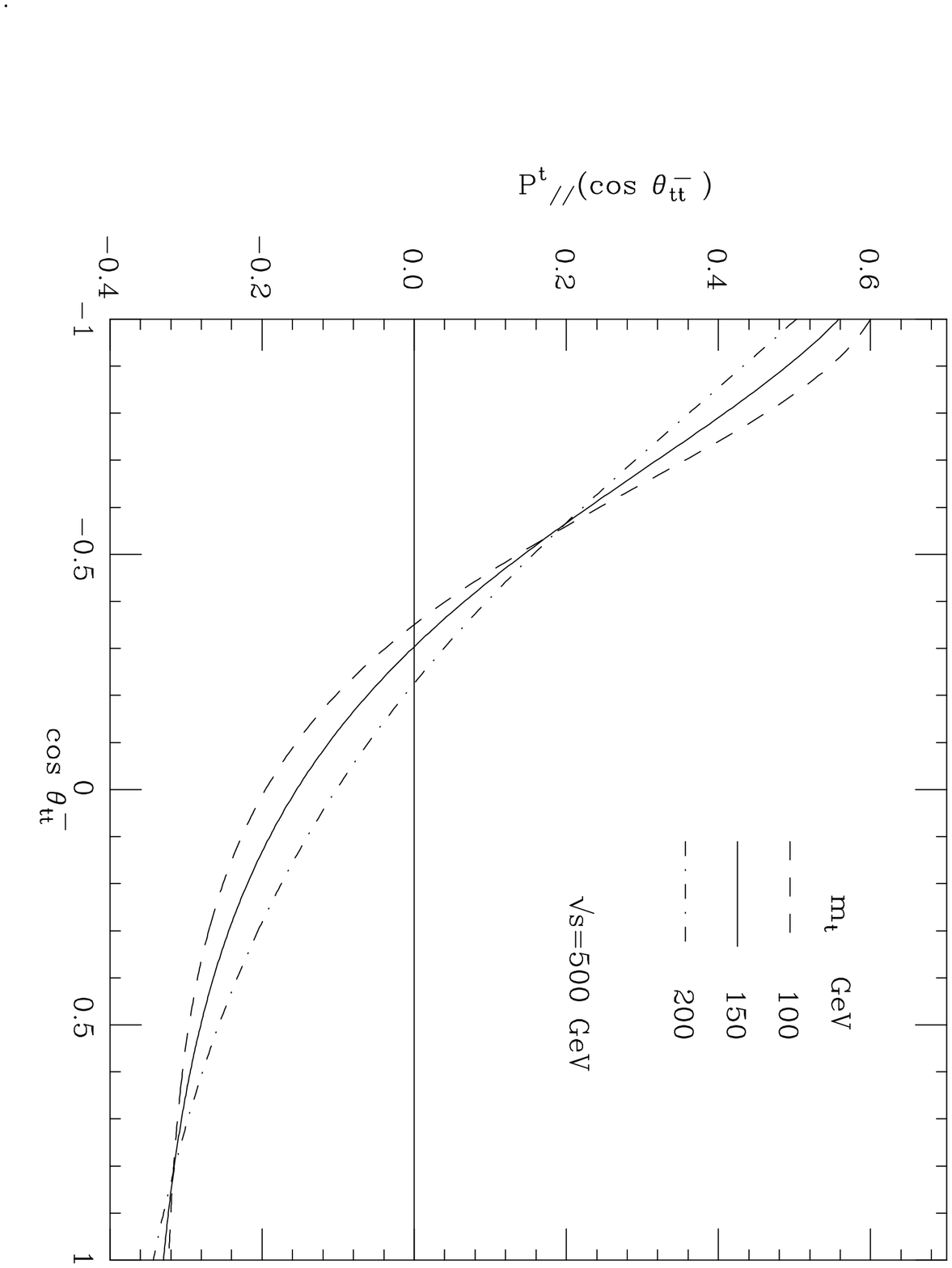,width=12cm}}
\end{sideways}
\end{center}
\caption{The longitudinal polarization of top quarks produced in
\ttprod\ , $P_\parallel^t$ as a function of the $\cos\theta_{\ttbar}$ for
 several top-quark masses, $m_t=100$, $150$ and $200$ GeV.}
\label{fig:polar}
\end{figure}

Using the above formulas $P_\parallel^t(\cos\theta_{\ttbar})$ can be
 calculated. We plot it in Fig.~\ref{fig:polar}.
\bigskip

The \as\ could also be rewritten as :
\begin{equation}
\as=-2h(m_t)\im({f_2^R}_{CPV})
\sum_{\tbhel=+,-}\left\{\frac{d\sigma(\ttprod)_{(+,\tbhel)}}
                             {d\sigma(\ttprod)}
                       -\frac{d\sigma(\ttprod)_{(-,\tbhel)}}
                             {d\sigma(\ttprod)}
\right\}.
\label{asymmfin}
\end{equation}
Here the subscripts $(\pm,h_{\bar t})$ specify the spin configuration
of the $t\bar t$ system.
In order to maximalize the number of available events we will present
also an integrated form of the asymmetry \as\ obtained by integration over
 \ttbar\ production phase space:
\begin{equation}
\as_{int}=-2h(m_t)\im({f_2^R}_{CPV})\frac{\int
 P_\parallel^t(\theta_{\ttbar})d\sigma
(\ttprod)}{\sigma(\ttprod)}.
\label{asymmetryint}
\end{equation}
As one can see from Fig.\ref{fig:polar},
$P_\parallel^t(\cos\theta_{\ttbar})$ switches
its sign between $0$ and $\pi$, therefore $\as_{int}$ integrated
over the full \ttbar\  phase space suffers from strong cancellation.
However, if we restrict the region of $\cos \theta_{\ttbar}$ in such a way
that the differential asymmetry \as\
is either positive or negative, we can effectively
enhance the integrated asymmetry.
\section{Illustration: Supersymmetric Standard Models}
In Ref.\cite{gg} the asymmetry \as\ has been calculated in the 2HDM,
where CP violation appears through mixing of CP-even and CP-odd states
in the  mass matrix of neutral scalars. CP violation within this model
has been  extensively discussed in the recent
literature~\cite{keung,gg,cp2hdm}. Many  papers have been devoted to
possible tests of CP violation induced in a SSM at
colliders~\cite{bg,susywerner,cpsusy}. Here we will calculate our \as\
also  within a SSM.
 
\begin{figure}[hbt]
\begin{center}\mbox{\epsfig{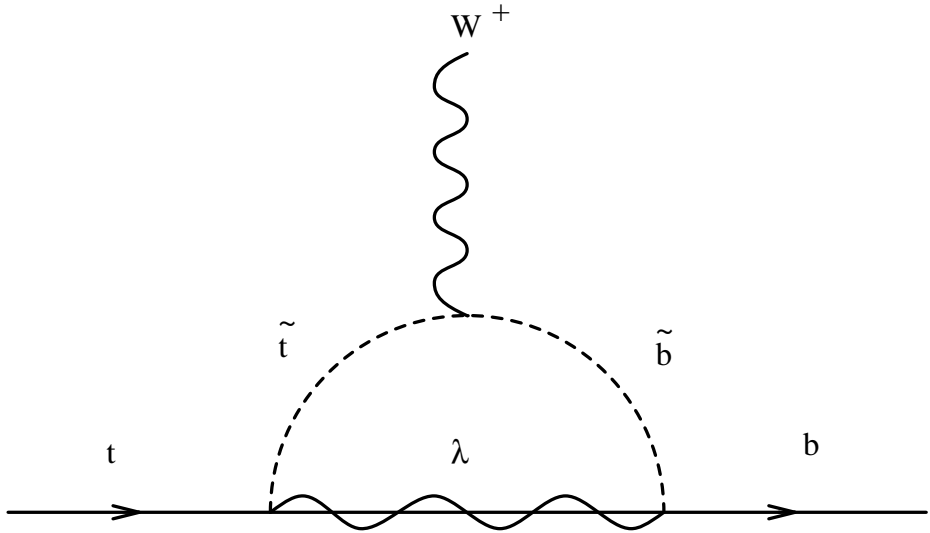}}\end{center}
\caption{The one-loop diagram, which generats $f_2^R$ within a SSM.}
\label{fig:diag}
\end{figure}

The asymmetry \as\ cannot be produced by any renormalizable
interactions at the tree level of perturbation expansion. Therefore it
may appear either as an artefact of some unknown high-scale theory in an
effective low-energy Lagrangian or it can be generated at a loop level
in the SM or its extensions. However, in the SM, ${f^R_2}_{CPV}$
vanishes, even at the one-loop level; we therefore illustrate the above
general consideration by the one-loop-generated ${f_2^R}_{CPV}$ in the
SSM.
We write down the relevant interaction:
\begin{eqnarray}
{\cal L} &=& i\surd 2 g_s
[{\tilde t}^\ast_L T^a (\bar {\lambda}^a t_L)+
{\tilde t}^\ast_R T^a (\bar {\lambda}^a t_R)] + (t\leftrightarrow b)
\nonumber\\
&&-(i/\sqrt{2})V_{tb}^{KM}
\tilde b_L^\dagger
\stackrel{\leftrightarrow}{\partial}_\mu
\tilde t_L W^{-\mu}
+ \hc,
\label{glulag}
\end{eqnarray}
where $g_s$ is the QCD coupling constant.
For our purpose the most relevant new source of CP violation, which
appears in the SSM, would be the phase in the $\tilde t_L$--$\tilde t_R$
mixing. The stop quarks of different handedness are related to the
stop-quark mass eigenstates ${\tilde t}_+$, ${\tilde t}_-$ through the
following transformation:
\begin{eqnarray}
\sstop_L & = & \cos \alpha_t \sstop_- - e^{ i\phi_t}\sin \alpha_t\sstop_+
\nonumber \\
\sstop_R & = & e^{-i\phi_t} \sin \alpha_t \sstop_- + \cos \alpha_t\sstop_+.
\label{mix}
\end{eqnarray}
The only one-loop diagram responsible for the generation of $f_2^R$ is
shown in Fig.~\ref{fig:diag}.
\bigskip

The $\tilde{b_L}$--$\tilde{b_R}$ mixing,
which may also provide the necessary phase, has the same structure as
the one for the top sector with the substitutions: $\phi_t \rightarrow
\phi_b$ and $\alpha_t \rightarrow \alpha_b$. However, if we assume
that the scalar $b$-quarks are almost degenerate, their mixing effect
can be neglected. The generalization is obvious.
It is interesting to note that, if we add phases $e^{i\phi_\lambda}$
and $e^{-i\phi_\lambda}$ to the terms $(\bar\lambda^a t_L)$ and
$(\bar\lambda^a t_R)$ in Eq.~(\ref{glulag}), owing to the complex gluino mass,
their effect can be absorbed into $\phi_t(\to \phi_t-2\phi_\lambda)$
and $\phi_b(\to \phi_b-2\phi_\lambda)$.
Since the same interactions generate the neutron's electric dipole moment
(NEDM), we have to take into account the limits originating  from this
measurement. However, direct restrictions on $\phi_{t/b}$ from the NEDM
turn out not to be very reliable~\cite{gavela} and therefore will not be
applied here. Indirect bounds may be obtained within the
supergravity-induced SSM. However, as showed in Ref.~\cite{kizukuri}, even assuming
the same phase for all quark families, the model allows for maximal
CP-violating phases for sufficiently heavy  up- and down-squarks;
therefore it is legitimate to assume maximal CP violation. It should be
stressed here that we are not restricting ourselves to the minimal
supergravity-induced models.

The result~\footnote{After completing this paper we have received a
preprint by  W. Bernreuther and P. Overmann~\cite{susywerner} where
$f_2^R$ has been also  calculated within SSM. We agree with their
results.} for the CP-violating part of  $f_2^R$  due to the
$\tilde{t_L}$--$\tilde{t_R}$ mixing can be written as:
\begin{equation}
\im({f_2^R}_{CPV})= \frac{\alpha_s}{3\pi}\sin(2\alpha_t) \sin(\phi_t)
\frac{m_\lambda m_W}{(m_t^2-m_W^2)^2}
\hbox{Re }\Bigl[{\cal I}(m^2_{\tilde t_+})-{\cal I}(m^2_{\tilde t_-})
          \Bigr]
\label{resform}
\end{equation}
\begin{eqnarray}
{\cal I}(m_{\tilde t}^2)&=&\int_0^1 \int_0^{1-\alpha}
{ d\beta d\alpha (m_t^2-m_W^2)^2 (1-\alpha-\beta) \over
\alpha m_{\tilde t}^2+\beta m_{\tilde b}^2
+(1-\alpha-\beta)m_\lambda^2+\alpha\beta(m_t^2-m_W^2)
-\alpha(1-\alpha)m_t^2}
\nonumber  \\
&=& (m_W^2-m_t^2)B_0(-p_b,\ml,\msb)+(m_W^2+m_t^2)B_0(-p_t,\ml,\mst )
 \nonumber \\
&\ &-2m_W^2B_0(p_W,\mst ,\msb)+C_0(-p_t,p_W,\ml,\mst ,\msb)\times  \\
&\ &
[(m_W^2-m_t^2)^2-(\mst -\ml-m_t^2)(m_W^2-m_t^2)-(\msb-\ml)(m_t^2+m_W^2)]
\nonumber
\end{eqnarray}
where $m_{{\tilde{t}}_{+/-}}$ stands for the top-squark masses,
$m_{\tilde{b}}$ denotes degenerated bottom squark mass and $m_\lambda$
is the gluino mass.
The functions $B_0(\cdots)$ and $C_0(\cdots)$ are defined as:
\begin{eqnarray}
&& \frac{i}{16\pi^2}B_0(k;m_1^2,m_2^2) \equiv \mu^{4-n} \int
\frac{d^n r}{(2\pi)^n} \frac{1}{[r^2-m_1^2+i\varepsilon]
[(r+k)^2-m_2^2+i\varepsilon]}  \\
&&\frac{-i}{16\pi^2}C_0(k,p;m_1^2,m_2^2,m_3^2) \equiv \\
&&\mu^{4-n} \int
\frac{d^n r}{(2\pi)^n} \frac{1}{[r^2-m_1^2+i\varepsilon]
[(r+k)^2-m_2^2+i\varepsilon]
[(r+k+p)^2-m_3^2+i\varepsilon]},\nonumber
\label{integ}
\end{eqnarray}
where $\mu$ is the regularization scale. $B_0$ and $C_0$ can be expressed in the standard manner through logarithms and Spence functions~\cite{velt}.
The result, Eq.~(\ref{resform}), is of course, UV finite.
In order to obtain effects as big as possible we will assume
for illustration \\ $\sin(2\alpha_t)\sin(\phi_t)=1$; for the lighter stop-quark
mass, we take the lower bound~\cite{stop} $m_{\tilde{t}_+}=50$~GeV, whereas for
the heavier one and for degenerated bottom squarks we use
$m_{\tilde{t}_+}=m_{\tilde{b}}=150$ GeV.
 
\begin{figure}[ht]
\begin{center}
\begin{sideways}
\mbox{\epsfig{file=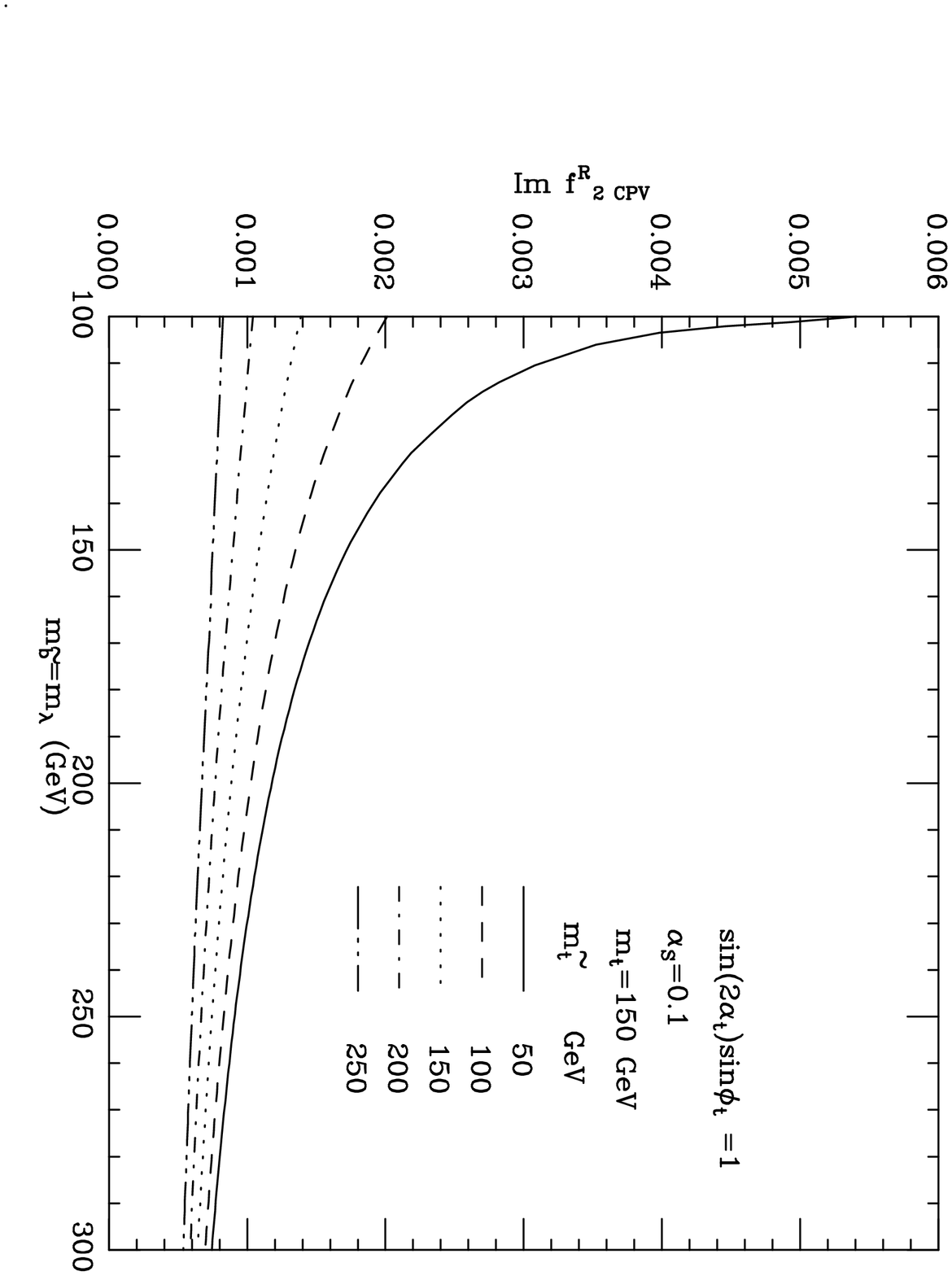,width=12cm}}
\end{sideways}
\end{center}
\caption{The CP-violating contribution to $\im(f_2^R)$ as a function of
$m_{\tilde{b}}=m_{\lambda}$.}
\label{fig:f2}
\end{figure}

Figure~\ref{fig:f2} shows
$\im({f_2^R}_{CPV})$, where we have fixed $\alpha_s=0.1$ and assumed
$m_{\tilde{t}_-}=\infty$.
Because of the unitarity form in
Eq.~(\ref{resform}), Im~$(f^R_{2\ CPV})$ can be obtained for the
general case of finite $m_{\tilde t_+}$ and $m_{\tilde t_-}$, by taking
the corresponding difference of two curves in Fig.~\ref{fig:f2}.
 
\begin{figure}[ht]
\begin{center}
\begin{sideways}
\mbox{\epsfig{file=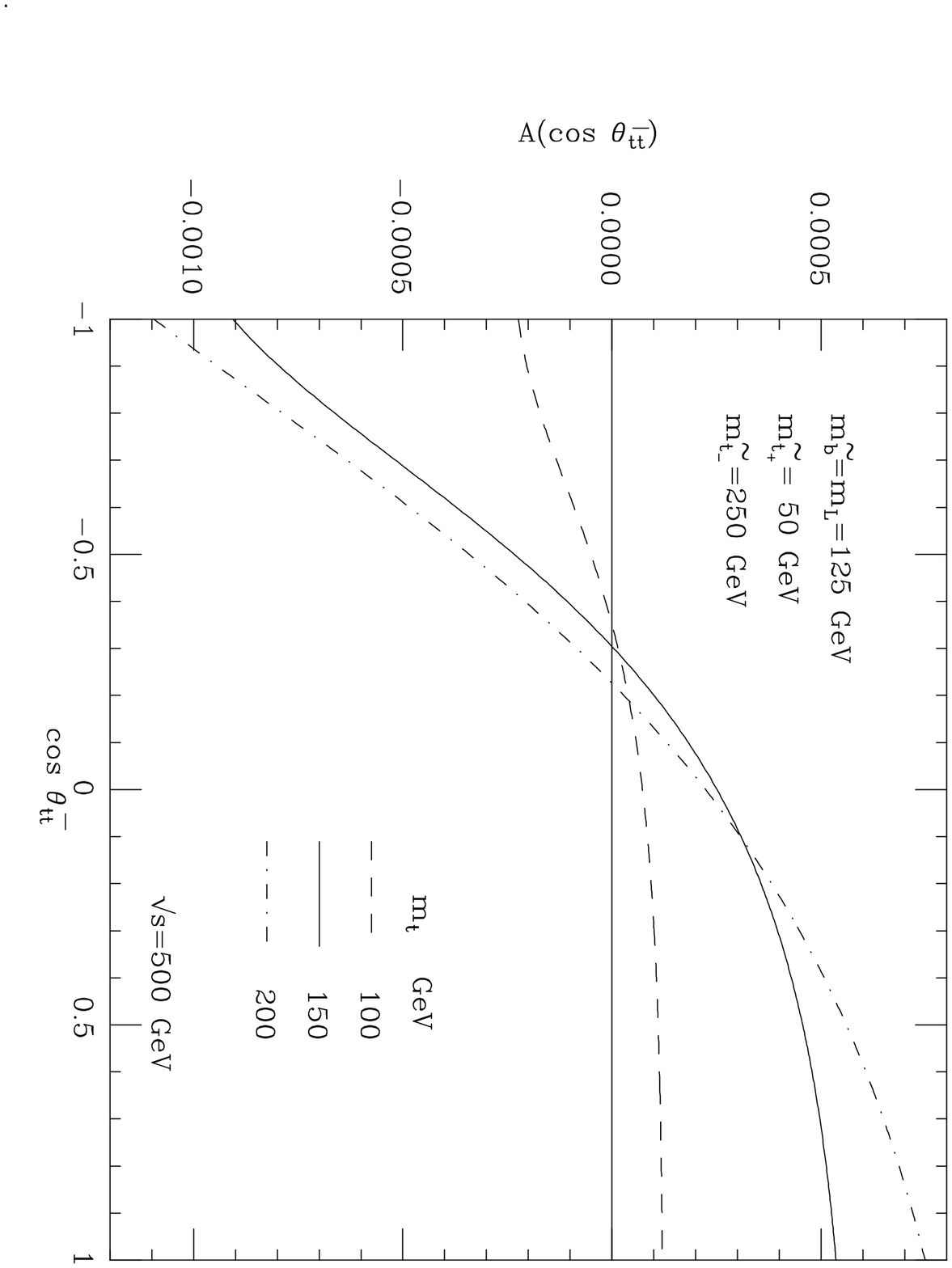,width=12cm}}
\end{sideways}
\end{center}
\caption{The asymmetry \as\ as a function of the $\cos\theta_{\ttbar}$ for
  top-quark masses $m_t=100$, $150$ and $200$ GeV.}
\label{fig:asdif}
\end{figure}

In Fig.~\ref{fig:asdif} we plot results for the asymmetry \as\ as a function of  $\cos \theta_{\ttbar}$.
 
\begin{figure}[ht]
\begin{center}
\begin{sideways}
\mbox{\epsfig{file=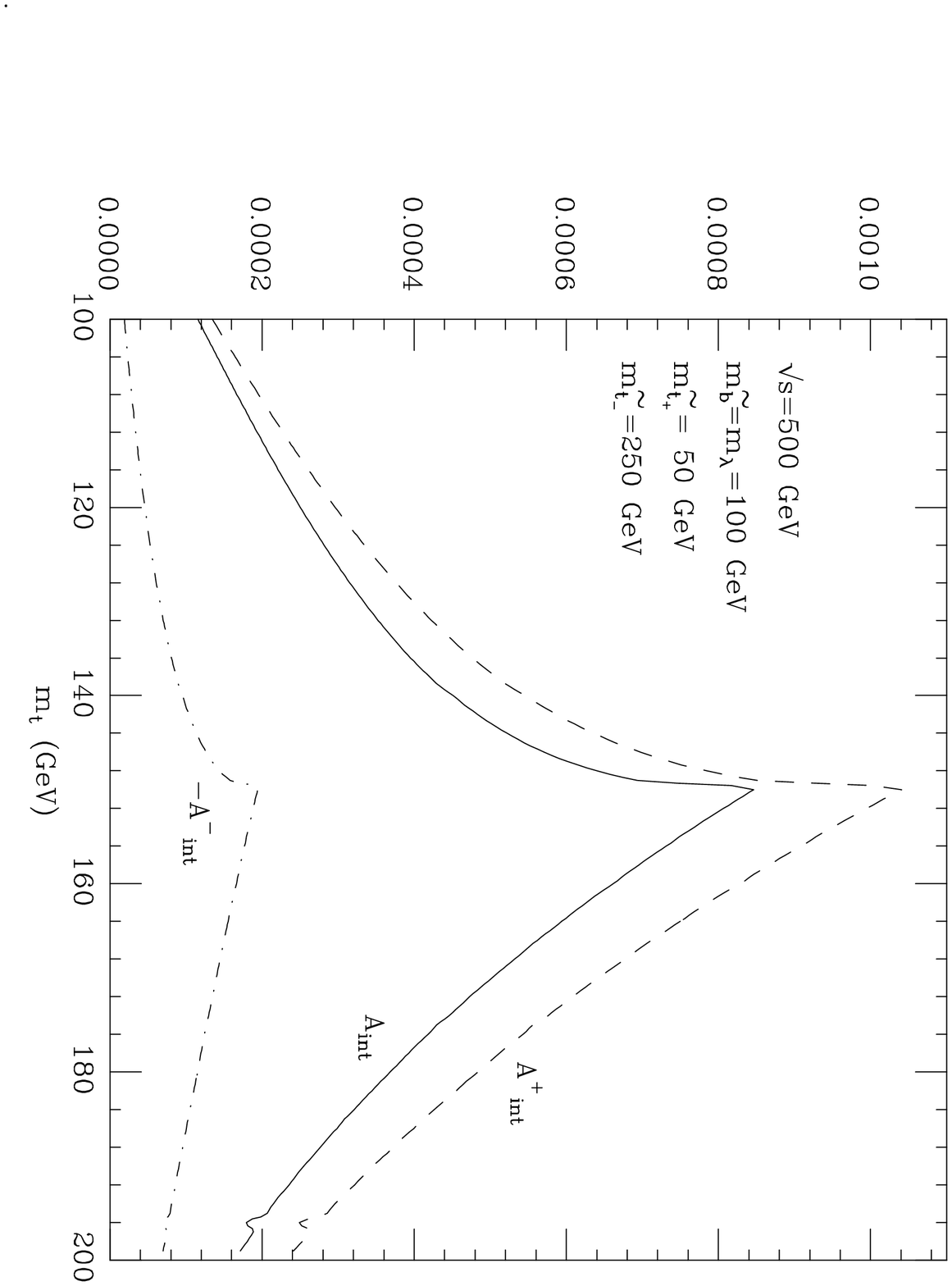,width=12cm}}
\end{sideways}
\end{center}
\caption{The integrated versions of the asymmetry \as\ ,  $\as_{int}^+$ and
 $\as_{int}^-$  defined in the text, as a function of the top quark mass.}
\label{fig:asint}
\end{figure}

In Fig.~\ref{fig:asint} we show
both $\as_{int}^+$ and $\as_{int}^-$ defined similarly as $\as_{int}$,
but with integration (in the numerator) over
$\cos \theta_{\ttbar}$ restricted to
regions of positive and negative differential asymmetry,
respectively.
\bigskip
\bigskip
\bigskip

For the most optimistic proposal of a $500\: \gev$
linear collider~\cite{topgroup},
the total integrated luminosity is $ 8.5 \times 10^4 \: \lumun $;
therefore the SM cross-section for $\ttbar$ production ($0.66\:\pbarn$
at $m_t=150\: \gev$)~\cite{topgroup} predicts about $N\simeq 50000$
events per year.
This means that the smallest possible measurable asymmetry is about
$1/\sqrt{B_\ell N} \simeq 1\%$. One should however
have in mind that the above
estimates do not include any cuts and, of course, some number of events
must be lost because of non-perfect efficiency.
The results that we have obtained here presumed the narrow-width
approximation, where all possible interference effects between
production and decay are neglected. In order to justify this
we must in addition
assume that the final $(Wb)$ mass resolution is sufficiently
good to be sure that $W$ and $b$ are coming from on-shell top quarks.
Since the largest asymmetry $\as_{int}$ is about
$10^{-4}\sim10^{-3}$, it seems very hard to find CP violation of the
$t$ decay in the SSM.
\section{Measurement of the Asymmetry}
 
\begin{figure}[ht]
\begin{center}
%\begin{sideways}
\mbox{\epsfig{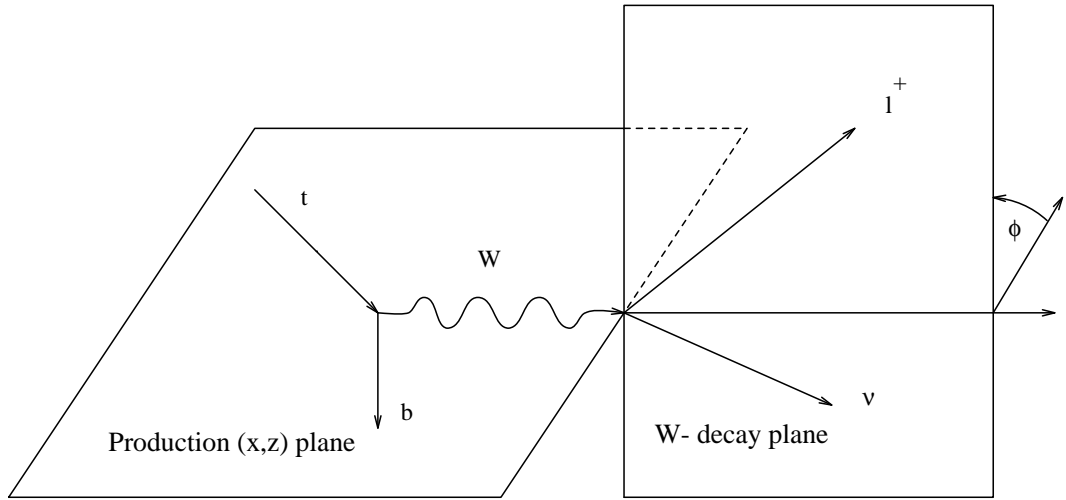}}
%\end{sideways}
\end{center}
\caption{The illustration of the azimuthal angle $\phi_{\ell^+}$.}
\label{fig:pic}
\end{figure}

Let us define
the 4-momentum of the $b$ quark and $W^+$ from the $t$ decay in the rest
frame of the $t$ quark as (we shall neglect $m_b$ in our kinematics):
\begin{eqnarray}
p_b^{t-rest}&=&E_b(1,-\sin\theta_{W^+},0,-\cos\theta_{W^+})\\
p_{W^+}^{t-rest}&=&(E_{W^+},E_b\sin\theta_{W^+},0,E_b\cos\theta_{W^+}),
\label{pbpw}
\end{eqnarray}
where $\theta_{W^+}$ is the polar angle defined in the top-quark rest frame,
 using for the $z$ axis the direction of the top seen from the $e^-e^+$
 centre-of-mass frame, the $(x,z)$ plane is defined by the top and $W^+$ momentum (having
 $p_{W^+} > 0$), and the $y$ axis is provided by the right hand rule.
(From simple kinematics we know that in the $t$ rest frame
$E_{W^+}=(m_t^2+m_W^2)/(2m_t)$
and $E_b=(m_t^2-m_W^2)/(2m_t)$.)
The axes of the $W^+$ rest frame
are defined by the sequence of transformations used to go from this to the $t$ rest frame; a detailed general description has been
 presented elsewhere~\cite{gg}. The choice of the $(x,z)$ plane in the top-quark
 centre of mass allows for a substantial simplification; it turns out that
 besides the obvious boost, the $W^+$ rest frame is obtained from the top-quark rest frame by a simple rotation about their common $y$ axis by the angle
 $\theta_{W^+}$. Therefore the $\phi_{\ell^+}$, which must be measured for the
 asymmetry determination, is simply the angle between the planes defined by
 $W^+-t$ and $W^+-\ell^+$, see Fig.~\ref{fig:pic}.
\bigskip

The easiest way to measure \as\ is offered by events where the top quark decays
semileptonically whereas the antitop decays purely hadronically.
In this case, in the $e^-e^+$  centre of mass (CM), the top quark momentum
$p_t^{CM}$ can be determined by
measurement of the antitop momentum. Since we need to know the $W^+$
momentum, we will assume hereafter that the $b$ and $\ell^+$ momentum are measured.
This is enough to reconstruct the $W^+$ rest frame and measure
$\phi_{\ell^+}$.
However, one can show the following useful formula:
\begin{equation}
{\bf p}_{\ell^+}^{CM}\cdot ({\bf p}_{b}^{CM}\times {\bf p}_{W^+}^{CM})
=E_{\ell^+}E_b\sin(\phi_{\ell^+})\sin(\theta_{W^+})
\sin(\theta_{\ell^+})\sinh(y_t)e^{y_{_
 W}},
\label{product}
\end{equation}
where $E_{\ell^+}=m_W/2$ is the $W^+$ rest frame $\ell^+$ energy,
and $y_t$, $y_W$ are the boost parameters for transformations from the CM frame
to the $t$ rest frame and from this frame to the $W$ rest frame,
respectively.
Since, $\theta_{W^+}$ and $\theta_{\ell^+}$ both lie between $0$ and $\pi$,
we find that the sign of $\sin\phi_{\ell^+}$ is given by the sign of
${\bf p}_{\ell^+}^{CM}\cdot ({\bf p}_{b}^{CM}\times {\bf p}_{W^+}^{CM})$.

\section{Summary}
We have discussed CP violation
in the decays $t \to \ell^+\nu b$ and $\bar t \to
\ell^-\bar\nu \bar b$ from the production $e^-e^+ \rightarrow \ttbar$.
The  asymmetry $\as_{int}$ defined by
Eq.~(\ref{asymmetryint}) turned out to be a  useful observable in
searching for non-standard sources of CP violation, since  it vanishes
even at the one-loop level in the SM. We can conclude that for the planned
luminosity at future high-energy linear colliders the observation of
$\as_{int}$ predicted within the SSM ($10^{-4}\sim10^{-3}$) looks
very difficult.
 
%\vspace{1cm}
%\centerline{\bf Acknowledgments}
%\vspace{.5cm}

\end{document}